\documentclass[conference]{IEEEtran}
\IEEEoverridecommandlockouts
\usepackage{cite}
\usepackage{comment}
\usepackage{amsmath,amssymb,amsfonts}
\usepackage{algorithmic}
\usepackage{graphicx}
\usepackage{textcomp}
\usepackage{multirow}
\usepackage{adjustbox}
\usepackage{makecell}
\usepackage{placeins}
\usepackage{hyperref} 
\usepackage{booktabs}
\usepackage{color}
\usepackage{lineno}
\usepackage{float}
\usepackage{bm}
\usepackage{soul}
\usepackage[table,xcdraw]{xcolor}
\usepackage{bigdelim}
\usepackage[table]{xcolor}
\usepackage{amssymb}

\newcommand{\bb}{\color{black}}

\newcommand{\bl}{\color{blue}}

\usepackage[numbers]{natbib}
\newcolumntype{?}{!{\vrule width 1.8pt}}

\def\BibTeX{{\rm B\kern-.05em{\sc i\kern-.025em b}\kern-.08em
    T\kern-.1667em\lower.7ex\hbox{E}\kern-.125emX}}
\begin{document}

\title{Texture-Aware StarGAN for CT data harmonization\\
}

\author{\IEEEauthorblockN{Francesco Di Feola\IEEEauthorrefmark{1}, Ludovica Pompilio \IEEEauthorrefmark{2}, Cecilia Assolito \IEEEauthorrefmark{3}, Valerio Guarrasi \IEEEauthorrefmark{4}, Paolo Soda \IEEEauthorrefmark{5}}

\IEEEauthorblockA{\IEEEauthorrefmark{2,4}Research Unit of Computer Systems and Bioinformatics, Campus Bio-Medico University of Rome, Rome, Italy}
\IEEEauthorblockA{\IEEEauthorrefmark{1,5}Department of Diagnostics and Intervention, Radiation Physics, Biomedical Engineering, Umeå University, Sweden}
\IEEEauthorblockA{\IEEEauthorrefmark{3} La Sapienza University of Rome}
\IEEEauthorblockA{\IEEEauthorrefmark{1}francesco.feola@umu.se
\IEEEauthorrefmark{2}ludovica.pompilio@alcampus.it 
\IEEEauthorrefmark{3} assolito.1857897@studenti.uniroma1.it
\IEEEauthorrefmark{4} 
valerio.guarrasi@unicampus.it \\
\IEEEauthorrefmark{5} paolo.soda@umu.se} }



\maketitle

\begin{abstract}
Computed Tomography (CT) plays a pivotal role in medical diagnosis; however, variability across reconstruction kernels hinders data-driven approaches, such as deep learning models, from achieving reliable and generalized performance.
To this end, CT data harmonization has emerged as a promising solution to minimize such non-biological variances
by standardizing data across different sources or conditions.
In this context, Generative Adversarial Networks (GANs) have proved to be a powerful framework for harmonization, framing it as a style-transfer problem.
However, GAN-based approaches still face limitations in capturing complex relationships within the images, which are essential for effective harmonization.
In this work, we propose a novel texture-aware StarGAN for CT data harmonization, enabling one-to-many translations across different reconstruction kernels.
Although the StarGAN model has been successfully applied in other domains, its potential for CT data harmonization remains unexplored. Furthermore, our approach introduces a multi-scale texture loss function that embeds texture information across different spatial and angular scales into the harmonization process, effectively addressing kernel-induced texture variations. We conducted extensive experimentation on a publicly available dataset, utilizing a total of 48667 chest CT slices from 197 patients distributed over three different reconstruction kernels, demonstrating the superiority of our method over the baseline StarGAN.
\end{abstract}


\section{Introduction}

Computed Tomography (CT) provides a powerful non-invasive visualization tool to reveal details about internal body
structures and serves
as a crucial decision support system for medical diagnosis \cite{hood2011predictive,aerts2014decoding}.
However, the inherent variability introduced by different reconstruction kernels poses significant challenges, especially for data-driven approaches such as deep learning systems, compromising their reliability and generalization, and limiting their clinical use.
To address these issues, CT data harmonization has emerged as a promising solution. The goal of CT data harmonization is to minimize non-biological variances, such as differences introduced by different reconstruction kernels \cite{seoni2024all}. 
This process enhances consistency across datasets, improving the reliability of data for analysis and facilitating more accurate and generalized model development.

To this end, although there exists a variety of harmonization techniques in the literature \cite{nan2022data}, whose review is out of the scope of this work, those based on generative modeling techniques have increasingly been adopted \cite{mali2021making, seoni2024all, nan2022data}.
These methods frame the harmonization task as a style transfer problem, where each group of data has a distinct style, and the samples have to be transferred to a target style.
Within this macro-category, paired methods rely on well-aligned datasets, where each image is available across all different styles. 
These methods typically employ algorithms based on convolutional neural networks (CNNs) \cite{park2019deep}, residual networks \cite{choe2019deep},  or Generative Adversarial Networks (GANs) \cite{selim2021stan, selim2021ct}.
The need for pairs represents a strong limitation for the actual application of these algorithms since the collection of coupled datasets is not only expensive and time-consuming, but is also unfeasible from a clinical perspective.
Due to the intrinsic unpaired nature of the existing datasets, unpaired methods are the most suitable to address the problem of CT data harmonization \cite{selim2022cross, kim2022multi, pei2023multi, selim2022uda}.
Among these, the CycleGAN framework
\cite{zhu2017unpaired} has emerged as a promising approach, offering cross-domain consistency through the cycle consistency constraint, thus eliminating the need for paired datasets.


However, the main limitation of CycleGAN-based approaches is that they only enable one-to-one translation, meaning that each model is trained to translate data from a specific source domain to a single target domain.
This constraint significantly limits their scalability and practicality for datasets with multiple source and target domains.
To this end, the StarGAN architecture \cite{choi2020stargan} emerges as a powerful alternative, enabling one-to-many translations within a single unified model.
The StarGAN framework has been successfully applied to various style transfer tasks, including facial attribute manipulation \cite{10194986} and artistic style conversion \cite{komatsu2021translation}.
In the medical domain, StarGAN has been employed for multi-site MRI harmonization \cite{liu2021style}; however, to the best of our knowledge, its potential for CT data harmonization remains unexplored.
Furthermore, existing GAN-based harmonization techniques  rely on standard loss functions, such as adversarial loss, reconstruction loss, and cycle-consistency losses \cite{pan2020loss}. 
While effective, these loss functions do not explicitly address textural details—an essential aspect in CT imaging, where reconstruction kernels significantly influence texture characteristics \cite{mayerhoefer2020introduction}.

On these grounds, we propose a novel texture-aware StarGAN for CT data harmonization that leverages a multi-scale texture loss function integrating texture information into its training process.  
By extracting texture details across different spatial and angular scales, our method effectively addresses the complex texture variations introduced by different reconstruction kernels.
We leverage the intrinsic multi-scale nature of the Gray-Level-Coocurence Matrix (GLCM), a handcrafted histogramming operation that counts how often different combinations of pixel intensity values occur in a specific spatial and angular relationship within an image. The GLCM implementation proposed here is a differentiable version that makes it compatible with gradient-based optimization.
Moreover, the process of embedding
texture descriptors into the harmonization framework is fully enabled by the use of a self-attention mechanism that provides for the dynamic end-to-end aggregation of the multiscale information.
The experimental evaluation, conducted on a publicly available dataset with three different reconstruction kernels, demonstrates the superiority of our approach compared to the baseline StarGAN model, underscoring the robustness of our method in addressing CT kernel harmonization.
Overall, our contributions can be summarized as follows:
\begin{itemize}
    \item We propose a novel texture-aware StarGAN designed for CT data harmonization, effectively addressing the challenge of kernel variability by embedding texture information into the harmonization process.
    \item We propose a dynamic aggregation rule that adaptively weights and integrates multi-scale  texture information into the model's loss function, enabling a more robust representation of texture.
    \item We perform extensive experimentation using a well-established publicly available dataset of chest CT scans across 3 reconstruction kernels, evaluating the effectiveness of harmonization by assessing alignment in both deep features and radiomic features.
    
\end{itemize}

The rest of the paper is organized as follows: section~\ref{sec:methods} outlines the methods, section~\ref{sec:ExperimentalConfiguration} describes the experimental configuration, section ~\ref{sec:Results} presents and discusses the results, and section~\ref{sec:Conclusion} provides concluding remarks.

\section{Methods} \label{sec:methods}
Let $K=\{k_1, k_2, ..., k_n\}$ be the set of CT reconstruction kernels, and let $\bm{x}_{k_i} \in \mathbb{R}^{h \times w}$ denote a CT scan acquired using kernel $k_i\in K$.
Given a target kernel $k_j\in K$ where $i \not= j$, image harmonization can be formulated as: 
\begin{equation}
    \bm{\hat{x}}_{k_{j}} = \phi(\bm{x}_{k_i}) \simeq \bm{x}_{k_{j}}
\end{equation}
where $\bm{\hat{x}}_{k_{j}}$ denotes the harmonized image, and $\phi$ is a generic harmonization function.

Our approach is illustrated in~\figurename~\ref{fig:method}.
Panel (a) presents the proposed texture-aware StarGAN framework, where the implemented StarGAN v2 \cite{choi2020stargan} consists of the following components:
a generator network $G$ that learns the harmonization function $\phi$; 
a mapping network $F$,  which generates style embeddings to condition the generation, enabling $G$ to perform one-to-many harmonization tasks; 
a style encoder $E$, which extracts style representation from input images; 
and a discriminator network $D$, necessary for adversarial training.

Texture awareness is enabled by two key components, embedded into the training of the StarGAN.
The first component is the Multi-Scale Texture Extractor (MSTE), detailed in panel (b), which derives a textural representation from calculating GLCMs at various spatial and angular scales. 
The second component is the aggregation module, detailed in panel (c), dynamically combining the textural representation into a scalar loss function $\mathcal{L}_{txt}$.
The resulting $\mathcal{L}_{txt}$ is combined with the baseline loss function, $\mathcal{L}_{baseline}$, which accounts for all the loss terms implemented in \cite{choi2020stargan}, effectively embedding textural information into the harmonization task.\\
In the following, we provide a mathematical description of how texture information is first extracted in Section \ref{lab:MSTE} and then embedded into the StarGAN framework in Section \ref{lab:aggregation}.

\subsection{Multi Scale Texture Extractor} \label{lab:MSTE}
\begin{figure*}[ht]
\centering
\includegraphics[width=\textwidth]{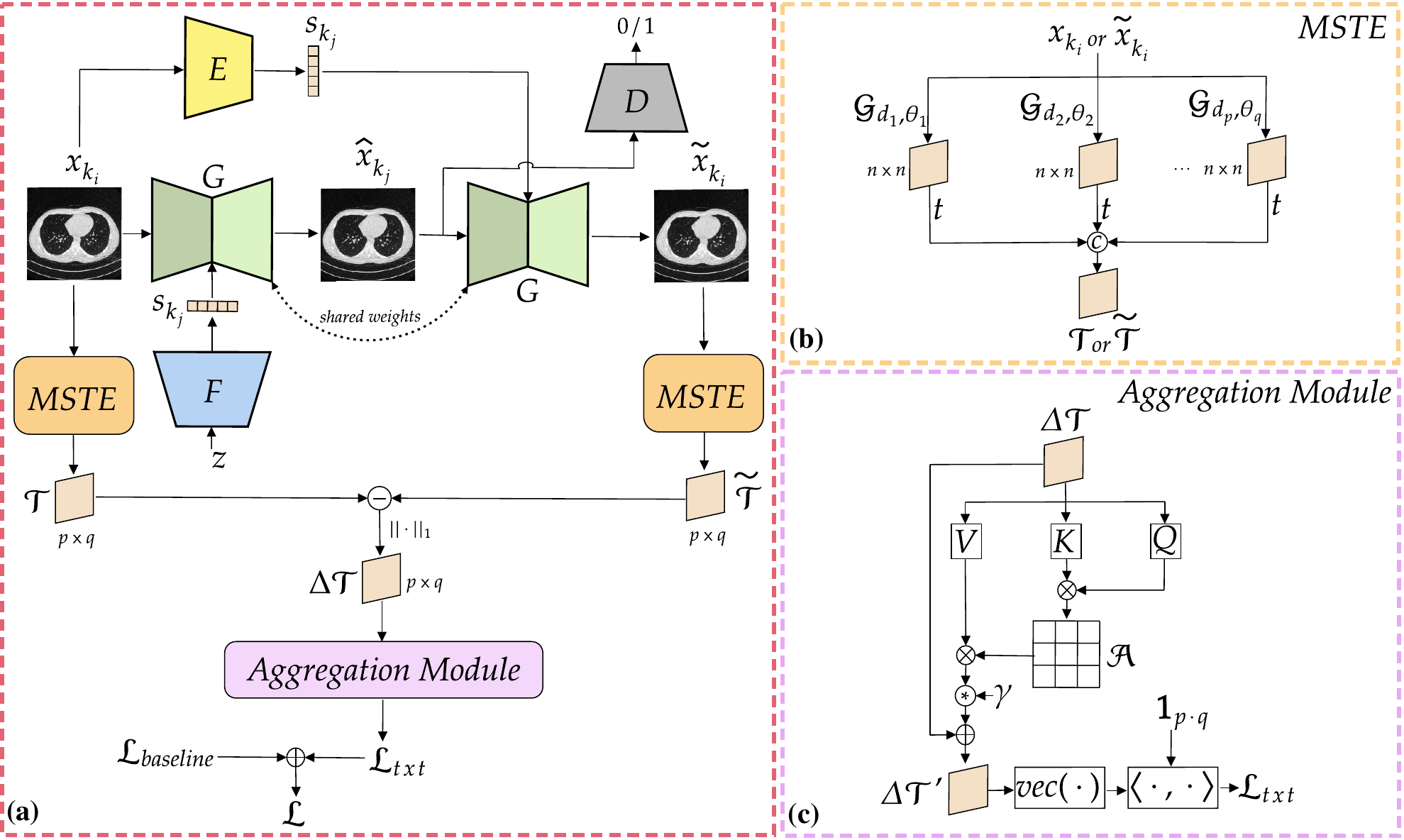}
\caption{Our texture-aware StarGAN for CT data harmonization. (a) Our framework includes a generator network $G$ that perform the harmonization, a mapping network $F$ generating style embeddings, a style encoder $E$, and a discriminator network $D$. Texture awareness is enabled by the Multi-Scale Texture Extractor (MSTE) and the aggregation module. (b) MSTE module extracting a textural representation ($\bm{\mathcal{T}}$ or $\bm{\Tilde{\mathcal{T}}}$ from the input image $\bm{x}_{k_i}$ or $\bm{\tilde{x}}_{k_j}$). (c) Dynamic aggregation that combines the extracted representation into a scalar loss function $\mathcal{L}_{txt}$.}
\label{fig:method}
\end{figure*}

Following~\figurename~\ref{fig:method} (a), the generator $G$ harmonizes the input image $\bm{x}_{k_i}$ as follows: 
\begin{equation}
    \bm{\hat{x}}_{k_j} = G(\bm{x}_{k_{i}}, \bm{s}_{k_{j}}) \simeq \bm{x}_{k_j},
\end{equation}
where $\bm{s}_{k_j}=F(\bm{z})$ is the style embedding synthesized by the mapping network $F$ from random noise $\bm{z}\sim N(0, I)$.
This embedding is used within the generator’s architecture, to perform Adaptive Instance Normalization (AdaIN) \cite{huang2017arbitrary},
which adjusts the mean and variance of the feature maps in the generator according to $\bm{s}_{k_j}$.
By encoding the characteristics of the target kernel $k_j$,
the embedding enables $G$ to transfer the desired style attributes to the input image $\bm{x}_{k_i}$.
The underlying principle is that image harmonization can be formulated as a style transfer process, aligning the statistical properties of the input with those of the target kernel.\\
Similarly, the style encoder $E$ extracts a style embedding from the input image, $\bm{s}_{k_j}=E(\bm{x}_{k_i})$, which is  used to map the generated image back to the original kernel, ensuring cycle-consistency:
\begin{equation}
    \bm{\tilde{x}}_{k_i} = G(\bm{\hat{x}}_{k_j}, \bm{s}_{k_i}).
\end{equation}

Then, both $\bm{\hat{x}}_{k_i}$ and $\bm{\tilde{x}}_{k_j}$ are passed to the MSTE module to extract a multiscale textural representation
based on GLCMs, i.e., $\bm{\mathcal{T}}$ and $\bm{\mathcal{\tilde{T}}}$.

We now describe how to extract $\bm{\mathcal{T}}$ and $\bm{\mathcal{\tilde{T}}}$. For fixed spatial and angular offsets denoted as $d$ and $\theta$, respectively, the GLCM is defined as a squared matrix $\bm{\mathcal{G}}_{d, \theta} \in \mathbb{R}^{n \times n}$ so that each element at coordinate $(i, j)$ lies in the range $[0,1]$.\\
Each of its elements is computed as:
\begin{equation}
 \bm{\mathcal{G}}_{d,\theta}(i,j) = \frac{g_{d, \theta}(i, j)}{\sum_{i=1}^{n}\sum_{j=1}^{n} g_{d,\theta}(i,j)}
\end{equation}
where,
\begin{equation}
 g_{d,\theta}(i,j) = \sum_{u=1}^{w}\sum_{v=1}^{h} \delta[ \bm{x}_{k_i}(u,v)=i]\cdot \delta [\bm{x}_{k_i}(\Tilde{u}, \Tilde{v})=j]
 \label{eq:glcm}
\end{equation}

s.t.
\begin{equation}
\Tilde{u} = u + d\cos\theta
\end{equation}
\begin{equation}
\Tilde{v} = v + d\sin\theta
\end{equation}
 and where $\delta$ is the Kronecker delta function, 
and $g_{d,\theta}(i,j)$ counts the occurrences of pixel value $i$ and $j$.

Notably, the GLCM is not differentiable by its very nature and it does not align with gradient-based optimization frameworks.
For this reason, we implemented it as a soft approximation which makes it a continuous and differentiable function. 
To this end, we employ a Gaussian soft assignment function for each pixel value to a set of predefined bins:
\begin{equation}
 a_k(\bm{x}_{k_i}(u,v)) = e^{-\frac{(\bm{x}_{k_i}(u,v) - b_k)^2}{2 \sigma^2}} \label{eq:soft}
\end{equation}
where $\bm{x}_{k_i}(u,v)$ is the pixel value in position $(u, v)$, $b_k$ is the $k^{th}$ bin value and $\sigma$ is a hyperparameter denoting the standard deviation of the Gaussian assignment function.

Let $P = \{d_{1}, d_{2},..., d_{p}\}$ and $Q = \{\theta_{1}, \theta_{2},..., \theta_{q}\}$ be the sets of spatial and angular offsets, respectively.
Let's now assume that $t$ is an operator extracting a texture descriptor from each $\bm{\mathcal{G}}_{d,\theta}$, defined as:
\begin{equation}
t(\bm{\mathcal{G}}_{d,\theta}) = \sum_{i=1}^{n}\sum_{j=1}^{n} f(i,j)\bm{\mathcal{G}}_{d,\theta}(i,j)
\end{equation}
where, in our implementation, $f(i, j)=(i-j)^2$.\\
We compute a multi-scale texture representation $\bm{\mathcal{T}} \in \mathbb{R}^{p\times  q}$ of $\bm{x}_{k_i}$ as:
\begin{equation}
\bm{\mathcal{T}} = 
\begin{bmatrix}
 t(\mathcal{G}_{d_{1},\theta_{1}}) & \cdots & \cdots & \cdots & t(\mathcal{G}_{d_{1},\theta_{q}})\\
 \vdots & \ddots &   & & \vdots \\
 \vdots &  & t(\mathcal{G}_{d_{i},\theta_{j}}) &  & \vdots \\
 \vdots &  &  & \ddots & \vdots \\
 t(\mathcal{G}_{d_{p},\theta_{1}}) & \cdots & \cdots & \cdots & t(\mathcal{G}_{d_{p},\theta_{q}})
\end{bmatrix}
\end{equation}
Similarly, for $\bm{\tilde{x}}_{k_i}$ we compute $\bm{\mathcal{\tilde{{T}}}} \in \mathbb{R}^{p\times v}$, as it is enough to replace $\bm{x}_{k_i}$ with $\bm{\tilde{x}}_{k_i}$ in Eq. \ref{eq:glcm}.

Going back to~\figurename~\ref{fig:method}, we then compute the error deviation $\Delta \bm{\mathcal{T}} \in \mathbb{R}^{p \times q}$ as:
\begin{equation}
 \Delta \bm{\mathcal{T}} = ||\bm{\mathcal{T}} - \bm{\mathcal{\tilde{T}}}||_{1}.
\end{equation}
\subsection{Aggregation Module} \label{lab:aggregation}
The aggregation module combines textural information into a multiscale texture loss function $\mathcal{L}_{txt}$.
The aggregation is dynamic since it enables the model to adaptively capture relationships between texture descriptors during the training of the model. 
Inspired by \cite{zhang2019self}, this approach is implemented as a self-attention layer
(Fig. \ref{fig:method} (c)): $\Delta \bm{\mathcal{T}}$ is passed through an attention layer which first applies $1 \times 1$ convolutions to extract keys $\bm{\mathcal{K}}$, queries $\bm{\mathcal{Q}}$ and values $\bm{\mathcal{V}}$.  
Then, the aggregation is computed as:
\begin{equation}
\mathcal{L}_{txt} = <\mathbf{1}_{p \cdot  q}, vec(\Delta \bm{\mathcal{T}}^{'})>
\end{equation}

where,
\begin{equation}
\Delta\bm{\mathcal{T}}^{'} = \gamma\bm{\mathcal{A}}\bm{\mathcal{V}} + \Delta\bm{\mathcal{T}}
\end{equation}
s.t.
\begin{equation}
\bm{\mathcal{A}} = SoftMax(\bm{\mathcal{Q}}^{T}\bm{\mathcal{K}}).
\end{equation}
$\Delta \bm{\mathcal{T}}^{'} \in \mathbb{R}^{p\times q}$ is the output of the attention layer, and $\mathbf{1}_{p \cdot q}$ is a vector of ones of size $p \cdot q$. The term $vec(\Delta \bm{\mathcal{T}}^{'})$ reshapes $\Delta \bm{\mathcal{T}}^{'}$ into a column vector, representing the vectorized error deviation. The inner product $<\cdot,\cdot>$  computes the weighted sum of all elements in $vec(\Delta \bm{\mathcal{T}}^{'})$, with uniform weights defined by $\mathbf{1}_{p \cdot q}$. $\bm{\mathcal{A}} \in \mathbb{R}^{(p \cdot q) \times (p \cdot  q)}$ denotes the attention map, and $\gamma$ is a trainable scalar weight.

Finally, as illustrated in~\figurename~\ref{fig:method} (a), after computing $\mathcal{L}_{txt}$, we combine it with the StarGAN's baseline loss function, $\mathcal{L}_{baseline}$, to form the overall loss $\mathcal{L}=\mathcal{L}_{baseline}+\mathcal{L}_{txt}$, effectively embedding multi-scale texture information into the harmonization task.

\section{Experimental evaluation} \label{sec:ExperimentalConfiguration}
This section outlines the experimental setup used to evaluate the proposed harmonization approach. Section \ref{materials} details the data used, Section \ref{performance} describes the quantitative metrics employed to assess the harmonization performance, while  Section \ref{implementation} provides a comprehensive overview of the training configurations and computational resources used.
\subsection{Materials}\label{materials}
We utilized a total of 48667 CT slices from 197 patients in the LIDC-IDRI dataset \cite{samuel2011lung}, a public repository of thoracic CT scans from patients diagnosed with lung cancer or those who underwent lung cancer screening. 
It was collected from seven centers and eight medical imaging companies, resulting in a diverse collection of CT scans acquired with different scanners and reconstruction kernels.
In this work, we grouped the data into three groups based on the reconstruction kernel, aiming to perform harmonization across CT scans within these groups.
The groups were defined as follows:
$k_1$ (Siemens Sensation 16-B30f) with  18790  slices from 60 patients, $k_2$ (GE Medical System LightSpeed QX-i-BONE) with 10355 slices from 81 patients, and $k_3$ (GE Medical Systems LightSpeed 16--STANDARD) with 19522 slices from 56 patients.
The rationale behind this grouping criterion is that the reconstruction kernel is a critical parameter in CT image acquisition and defines how raw data is processed into interpretable images. Different reconstruction kernels emphasize various features, such as sharpness, noise levels, and texture, directly influencing the radiomics features extracted from the images \cite{mayerhoefer2020introduction}. These variations can introduce significant heterogeneity across datasets, making it challenging to derive consistent and comparable quantitative measures. 

To ensure unbiased model development and evaluation, the dataset was split into training and test sets in an 80/20 hold-out approach.
In this work, we opted to work with 2D slices instead of 3D volumes for two reasons: first, the number of patients was not enough to train
a model working with 3D data; second, while being aware that a 3D approach would catch more complete spatial relationships, the 2D approach is the most used in the literature, and so we
deemed it was best suited for the scope of this work.

We preprocessed all the images as follows:
first, we converted all the raw DICOM files to Hounsfield Units (HU). 
After resampling to a uniform spacing of $[1, 1, 1]$, we cropped a region of interest containing the whole anatomical structure of the lungs and centered it within a $512\times512$ image, with the background value set to $-$1024.
Finally, the pixel values were normalized in $[-1, 1]$ for subsequent processing.

\subsection{Performance evaluation}\label{performance}
The performance evaluation involved two complementary analyses: deep features and radiomic features.
This dual approach was chosen to provide a comprehensive understanding of the harmonization task: on the one side, deep features, extracted using deep learning models, capture high-level representations sensitive to global and structural variations.
On the other side, radiomic features focus on specific characteristics, such as texture patterns and pixel intensities, providing a detailed analysis of local variations.

\subsubsection{Deep feature analysis}
We used the well-established Fréchet Inception Distance (FID) \cite{heusel2017gans} for deep feature analysis.
FID measures the similarity between the feature distributions of real and generated images, using a pretrained InceptionV3 \cite{szegedy2016rethinking} as the feature extractor.
It is computed as:
\begin{equation}
 FID = \|\mu_r - \mu_g\|^2 + \operatorname{Tr}(\Sigma_r + \Sigma_g - 2(\Sigma_r \Sigma_g)^{1/2}).
\end{equation}
where $\operatorname{Tr}(\cdot)$ denotes the trace operator, $\mu_r$, $\Sigma_r$, and $\mu_g$, $\Sigma_g$ represent the mean and covariance of real and generated images, respectively. 
A lower FID score indicates that the generated images closely resemble the real images in terms of feature distribution.

\subsubsection{Radiomics features}\label{radiomics_workflow}
To analyze handcrafted features, we extracted a set of 102  radiomic features, denoted as $\mathcal{P}$, using the Pyradiomics library \cite{van2017computational}.
They are grouped into the following classes:
\begin{itemize}
    \item First-order statistics (19 features).
    \item Gray-Level-Co-occurrence Matrix (24 features).
    \item Gray-Level-Run-Length-Matrix (16 features).
    \item Gray-Level-Size-Zone-Matrix (16 features).
    \item Gray-Level-Dependence-Matrix (14 features).
    \item Neighboring Gray Tone difference Matrix (5 features).
\end{itemize}
To evaluate the harmonization process, we followed the structured workflow outlined below: 
\\
\emph{Before harmonization}, we extracted all radiomic features from the test set.
For each feature and each pair of reconstruction kernels ($k_1\leftrightarrow{}k_2$, $k_2\leftrightarrow{}k_3$, and $k_1\leftrightarrow{}k_3$),  we applied the Welch statistical test to identify the subset of features, $\mathcal{R} \subseteq \mathcal{P}$, that showed statistically significant differences in their distributions between the two domains. 
In others words, we identified features that were statistically different across each pair of reconstruction kernels.
A p-value threshold of $<0.01$ was used for this analysis.\\
\emph{After harmonization}, we re-extracted  all features from the isolated subgroup $\mathcal{R}$ from the harmonized images and compared them to the features derived from real images natively reconstructed with the target kernel. 
Then, using the Welch statistical test, we identified the subset of features, $\mathcal{Z} \subseteq \mathcal{R} \subseteq \mathcal{P}$,  that no longer showed statistically significant differences between the two distributions.
This result indicates that the harmonization process effectively aligned the radiomic feature distributions in $\mathcal{Z}$ with those of real images, demonstrating its effectiveness in standardizing radiomic features.


\subsection{Implementation details}   \label{implementation}
The computational setup utilized a high-performance computing cluster equipped with four NVIDIA V100 GPUs, each with 40 GB of memory, to accommodate the intensive computational requirements of GAN training.
Each generator and discriminator was initialized with random weights and trained from scratch. 
To ensure consistency across experiments, the same network architectures, hyperparameters, and training protocols were applied to all models trained, using the default parameters specified in the original implementation of the StarGAN v2 \cite{choi2020stargan}.
We used a set of spatial offsets $P = \{1, 3, 5, 7\}$ and a set of angular offsets $Q = \{0^{\circ}, 45^{\circ}, 90^{\circ}, 135^{\circ}\}$.
We trained both models for up to 100000 iterations, with a batch size of 8 and a learning rate equal to $10^{-4}$.

\section{Results and discussion} \label{sec:Results}
This section performs an in-depth analysis to assess our texture-aware StarGAN for CT data harmonization, providing both quantitative analysis and visual examples. Specifically, we analyses deep features in section \ref{lab:fid}, and radiomic features in section \ref{lab:radiomics}.

\subsection{Deep feature analysis} \label{lab:fid}
\tablename~\ref{table:1} presents the FID scores for both the baseline StarGAN model and our proposed texture-aware StarGAN, comparing results before and after harmonization at the end of training.
Each row corresponds to a pair of reconstruction kernels, with the symbol $\leftrightarrow$ denoting bidirectional harmonization. 
The numerical results are averaged over the two harmonization directions, $k_i\rightarrow{}k_j$ and $k_j\rightarrow{}k_i$, with $i\not=j$, ensuring a balanced evaluation of harmonization performance. 
We employ bold text for the best results.
The results highlight the clear advantage of our texture-aware StarGAN over the baseline StarGAN. For most kernel translation pairs, our approach consistently achieves lower FID scores compared to the baseline.
However, it is important to note that for the  $k_1\leftrightarrow{}k_3$ pair, which shows the highest FID score before harmonization,  our texture-aware StarGAN results in a slightly higher FID score compared to the baseline.
This outcome highlights the complexity of aligning domains with substantial differences in global and structural characteristics.

\begin{figure}[ht]
\centering
\includegraphics[width=0.485\textwidth]{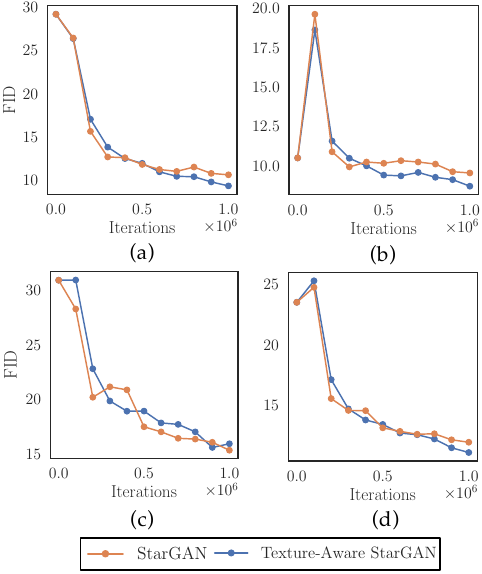}
\caption{Evolution of FID score during training: (a) $k_1\leftrightarrow{}k_2$, (b) $k_2 \leftrightarrow{}k_3$, (c)  $k_1\leftrightarrow{}k_3$, and (d) the average trend across all harmonizations performed by the models. The symbol $\leftrightarrow{}$ denotes bidirectional harmonization.
}
\label{fig:fid_trend}
\end{figure}

To gain further insight into the harmonization process, \figurename~\ref{fig:fid_trend} presents a detailed visualization of the FID score convergence patterns during training for each kernel translation pair in \figurename~\ref{fig:fid_trend}  (a)-(b)-(c) and the average trend across all harmonizations performed by the models in \figurename~\ref{fig:fid_trend} (d).
In \figurename~\ref{fig:fid_trend} (a), the FID scores for both harmonization methods decrease as training progresses, indicating improved alignment between source and target reconstruction kernel distributions.
However, texture-aware StarGAN converges more quickly and achieves a noticeably lower FID value. In \figurename~\ref{fig:fid_trend} (b), aside from the initial peak—likely due to training instability—a similar trend is observed:  the baseline StarGAN shows a slower decline in FID and stabilizes at a higher value, while the proposed method shows a  sharper reduction throughout the entire training process.
In \figurename~\ref{fig:fid_trend} (c),  the most challenging pair due to higher initial FID, 
the convergence rates of both approaches are comparable.
Although the final FID  of the proposed method is slightly higher than that of the baseline, our texture-aware StarGAN shows a smoother decrease in FID over time.
Moreover, the average performance trend across all the harmonizations, shown in \figurename~\ref{fig:fid_trend} (d), supports the effectiveness of texture awareness in the harmonization process.
Notably, the proposed method demonstrates a faster convergence and lower average FID values compared to the baseline StarGAN.

\begin{table}[h!]
    \centering
    \caption{Frechet Inception Distance (FID) before and after harmonization measured after $10^5$ training iterations. The symbol $\leftrightarrow{}$ indicates bidirectional harmonization.
    }
    \begin{adjustbox}{width=\columnwidth}
    \begin{tabular}{|c|c|c|c|}
        \hline
        \multirow{2}{*}{\textbf{Kernels}} & \multicolumn{3}{c|}{\textbf{ FID $\downarrow$}} \\
        \cline{2-4}
         & \textbf{Before Harmonization} & \textbf{StarGAN} &\textbf{Texture-Aware StarGAN}\\ 
        \hline
         $k_1 \leftrightarrow{} k_2$ & 29.16 & 10.63 & \textbf{9.34} \\

        \hline
        $k_2 \leftrightarrow{} k_3$ & 10.49 & 9.54 & \textbf{8.69} \\
        \hline
        $k_1 \leftrightarrow{} k_3$ & 30.90 & \textbf{15.31} & 15.90 \\
        \hline

    \end{tabular}
    \end{adjustbox}
    \label{table:1}
\end{table}

\subsection{Radiomics features} \label{lab:radiomics}
\begin{figure}[ht]
\centering
\includegraphics[width=0.47\textwidth]{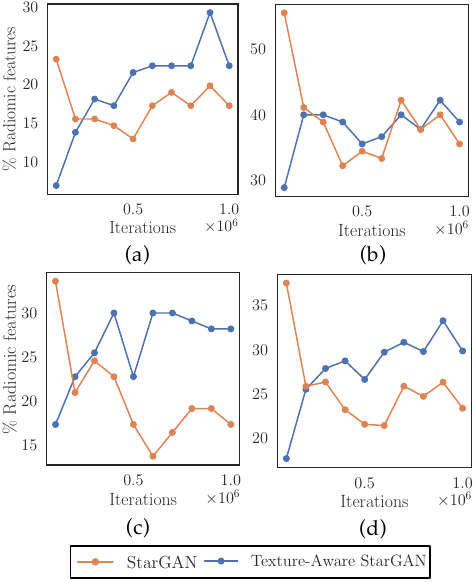}
\caption{Evolution of the percentage of radiomics features with no statistical difference after harmonization: (a) $k_1 \leftrightarrow k_2$, (b) $k_2 \leftrightarrow k_3$, (c) $k_1 \leftrightarrow k_3$, and (d) average trend across all harmonizations performed by the models. The symbol $\leftrightarrow$ denotes bidirectional harmonization.}
\label{fig:radiomics_trend}
\end{figure}

\begin{figure}[H]
\centering
\includegraphics[width=0.5\textwidth]{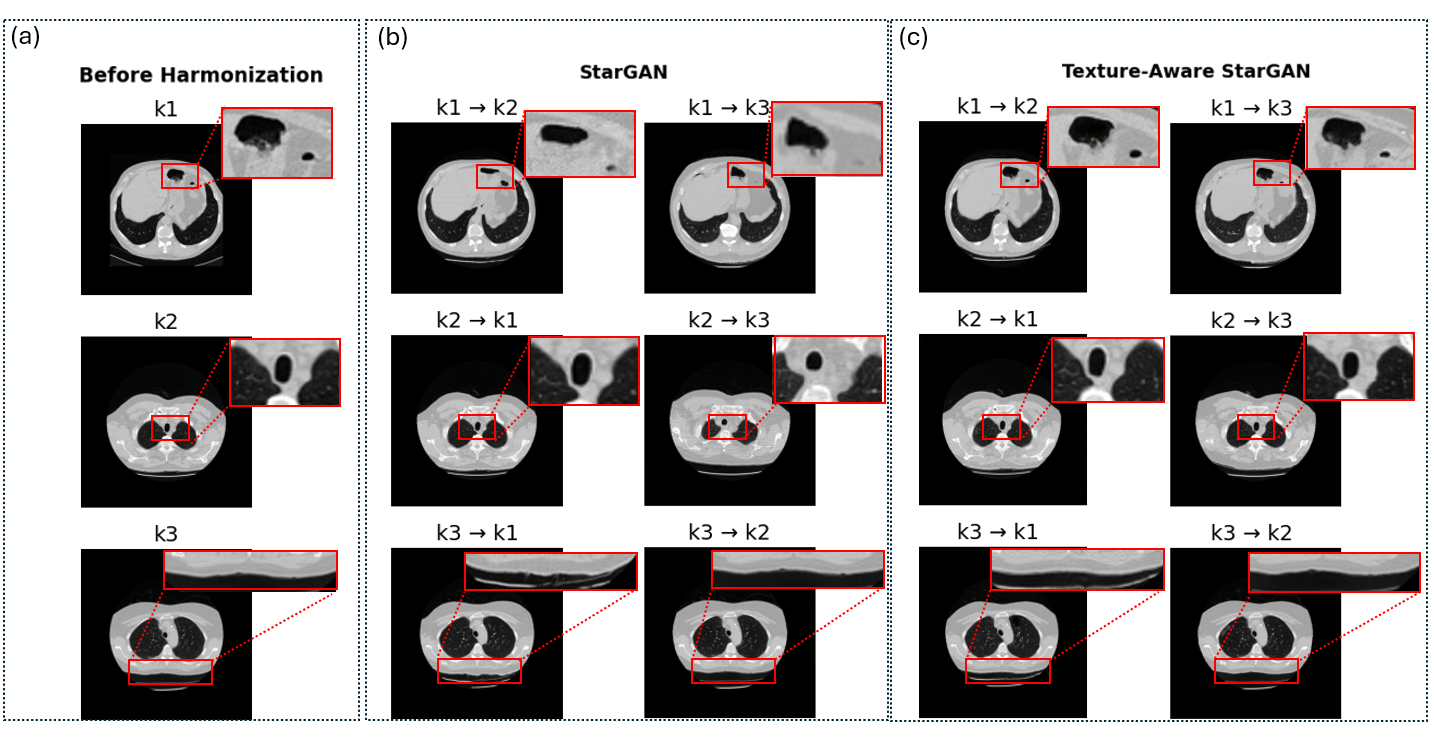}
\caption{Visual comparison of CT images before harmonization and after harmonization using the baseline StarGAN and the proposed texture-aware StarGAN.}
\label{fig:visual}
\end{figure}

\tablename~\ref{table:2} shows the alignment between radiomics features after harmonization for both the baseline StarGAN and the proposed texture-aware StarGAN at the end of the training, computed according to the workflow outlined in section \ref{radiomics_workflow}.
Each row corresponds to a pair of reconstruction kernels, with the symbol $\leftrightarrow$
denoting bidirectional harmonization as before. 
For all reconstruction kernel pairs, the results show a marked improvement in feature alignment when using our texture-aware StarGAN compared to the baseline StarGAN.
Notably, for $k_1 \leftrightarrow{} k_2$, the baseline StarGAN achieves an average alignment
of 17.24\% of the features, while the texture-aware StarGAN raises this value to 22.42\%.
For $k_2 \leftrightarrow k_3$, the alignment percentages are 35.55\% for the baseline StarGAN and 38.89\% for the texture-aware StarGAN.
While the improvement is less pronounced compared to $k_1 \leftrightarrow k_2$, these results indicate that texture awareness reduces discrepancies, even between kernel pairs with more similar structural properties (see the first column of \tablename~\ref{table:1}).
For $k_1 \leftrightarrow k_3$ the baseline StarGAN reaches a feature alignment of 17.27\% whilst the proposed texture-aware StarGAN, raises this value to 28.19\%, highlighting its robustness even in harmonizing challenging kernel pairs with larger differences in terms of deep features (see the first column of Table \ref{table:1}).
To gain further insight into the harmonization process, \figurename~\ref{fig:radiomics_trend} presents the evolution of feature alignment over training for each kernel translation pair in \figurename ~\ref{fig:radiomics_trend} (a)-(b)-(c) and the average trend across all harmonizations performed by the models 
\figurename~\ref{fig:radiomics_trend} (d).
In \figurename~\ref{fig:radiomics_trend} (a), our texture-aware StarGAN exhibits a consistent upward trend, rapidly surpassing the baseline StarGAN and maintaining a higher percentage of aligned features as training progresses.
In \figurename~\ref{fig:radiomics_trend} (b), the trends show slower and more variable improvement for both methods;
however, our texture-aware StarGAN still achieves a higher percentage of aligned features. 
In \figurename~\ref{fig:radiomics_trend} (c) the gap between the two harmonization approaches is even more pronounced, as the baseline StarGAN shows decreasing feature alignment throughout training, whereas the texture-aware StarGAN maintains a consistent upward trend.
The overall trend shown in \figurename~ \ref{fig:radiomics_trend} (d) strongly supports our findings, underscoring  the consistent effectiveness of the proposed method in aligning textural characteristics across all kernel translation pairs, highlighting the critical role of texture awareness in enhancing the harmonization process throughout model training.
Finally, \figurename~\ref{fig:visual} shows an example of harmonized CT images.
Panel (a) presents the images before harmonization for the three reconstruction kernels, i.e., $k_1$, $k_2$ and $k_3$.
Panel (b) shows the harmonized images using the baseline StarGAN model, whereas panel (c) reports the harmonized images produced by our texture-aware StarGAN.
To facilitate the reader, we included zoomed-in Region of Interest (ROIs), marked with red boxes, capturing critical structural details within the images.
A visual inspection highlights that structural details are better preserved by the harmonization approach proposed here, as emphasized by the zoomed-in ROIs.

\begin{table}[h!]
    \centering
    \caption{Percentage of radiomics features with no statistical difference after harmonization. The symbol $\leftrightarrow{}$ denotes bidirectional harmonization.
    }
    \begin{adjustbox}{width=0.7\columnwidth}
    \begin{tabular}{|c|c|c|}
        \hline
        \multirow{2}{*}{\textbf{Kernels}} & \multicolumn{2}{c|}{\textbf{\% of radiomics features} $\uparrow$} \\
        \cline{2-3}
   & \textbf{StarGAN} &\textbf{Texture-Aware StarGAN}\\ 
        \hline
         $k_1 \leftrightarrow{} k_2$  &  17.24 & \textbf{22.42} \\

        \hline
        $k_2 \leftrightarrow{} k_3$  & 35.55 & \textbf{38.89} \\
        \hline
        $k_1 \leftrightarrow{} k_3$  & 17.27 & \textbf{28.19} \\
        \hline

    \end{tabular}
    \end{adjustbox}
    \label{table:2}
\end{table}

\section{Conclusion} \label{sec:Conclusion}
In this work, we have proposed a texture-aware StarGAN for CT data harmonization.
The core methodological contributions are threefold:
(1) we explore the use of the StarGAN model for CT data harmonization, which, to the best of our knowledge, has not been previously investigated; (2) we introduce texture awareness by embedding texture information into the harmonization process, leveraging texture descriptors captured across different spatial and angular scales and dynamically integrating them into the model's loss function; and (3) we validate our approach on a public dataset with more than 45000 CT slices spanning 3 CT reconstruction kernels.
Our comprehensive analysis focused on the alignment of both deep features, measured by the FID score, and radiomics features. 
Although we did not observe a large improvement in deep feature alignment with our texture-aware StarGAN compared to the baseline StarGAN, we found a considerable enhancement in radiomic features alignment.
These results underscore the critical role of introducing texture-awareness into the harmonization pipeline.
Further analysis should be carried out to investigate the scalability of our approach to other datasets with a broader range of reconstruction kernels and to evaluate its potential as a preprocessing step in downstream deep learning pipelines, as well as its utility in radiology workflow.
Indeed, an effective harmonization tool could provide radiologists with more consistent and reliable CT image assessments.
Additionally, future work will investigate hybrid approaches for CT data harmonization that will combine the strengths of GANs with the latest advancements in diffusion-based modeling.

\section{Acknowledgments}

\bibliography{biblio.bib}

\begin{thebibliography}{10}
\providecommand{\url}[1]{#1}
\csname url@samestyle\endcsname
\providecommand{\newblock}{\relax}
\providecommand{\bibinfo}[2]{#2}
\providecommand{\BIBentrySTDinterwordspacing}{\spaceskip=0pt\relax}
\providecommand{\BIBentryALTinterwordstretchfactor}{4}
\providecommand{\BIBentryALTinterwordspacing}{\spaceskip=\fontdimen2\font plus
\BIBentryALTinterwordstretchfactor\fontdimen3\font minus \fontdimen4\font\relax}
\providecommand{\BIBforeignlanguage}[2]{{%
\expandafter\ifx\csname l@#1\endcsname\relax
\typeout{** WARNING: IEEEtran.bst: No hyphenation pattern has been}%
\typeout{** loaded for the language `#1'. Using the pattern for}%
\typeout{** the default language instead.}%
\else
\language=\csname l@#1\endcsname
\fi
#2}}
\providecommand{\BIBdecl}{\relax}
\BIBdecl

\bibitem{hood2011predictive}
L.~Hood and S.~H. Friend, ``{Predictive, personalized, preventive, participatory (P4) cancer medicine},'' \emph{Nature reviews Clinical oncology}, vol.~8, no.~3, pp. 184--187, 2011.

\bibitem{aerts2014decoding}
H.~J. Aerts, E.~R. Velazquez, R.~T. Leijenaar, C.~Parmar, P.~Grossmann, S.~Carvalho, J.~Bussink, R.~Monshouwer, B.~Haibe-Kains, D.~Rietveld \emph{et~al.}, ``{Decoding tumour phenotype by noninvasive imaging using a quantitative radiomics approach},'' \emph{Nature communications}, vol.~5, no.~1, p. 4006, 2014.

\bibitem{seoni2024all}
S.~Seoni, A.~Shahini, K.~M. Meiburger, F.~Marzola, G.~Rotunno, U.~R. Acharya, F.~Molinari, and M.~Salvi, ``{All you need is data preparation: A systematic review of image harmonization techniques in Multi-center/device studies for medical support systems},'' \emph{Computer Methods and Programs in Biomedicine}, p. 108200, 2024.

\bibitem{nan2022data}
Yang, J.~Del~Ser, S.~Walsh, C.~Sch{\"o}nlieb, M.~Roberts, I.~Selby, K.~Howard, J.~Owen, J.~Neville, J.~Guiot \emph{et~al.}, ``{Data harmonisation for information fusion in digital healthcare: A state-of-the-art systematic review, meta-analysis and future research directions},'' \emph{Information Fusion}, vol.~82, pp. 99--122, 2022.

\bibitem{mali2021making}
S.~A. Mali, A.~Ibrahim, H.~C. Woodruff, V.~Andrearczyk, H.~M{\"u}ller, S.~Primakov, Z.~Salahuddin, A.~Chatterjee, and P.~Lambin, ``{Making radiomics more reproducible across scanner and imaging protocol variations: a review of harmonization methods},'' \emph{Journal of personalized medicine}, vol.~11, no.~9, p. 842, 2021.

\bibitem{park2019deep}
S.~Park, S.~M. Lee, K.-H. Do, J.-G. Lee, W.~Bae, H.~Park, K.-H. Jung, and J.~B. Seo, ``{Deep learning algorithm for reducing CT slice thickness: effect on reproducibility of radiomic features in lung cancer},'' \emph{Korean journal of radiology}, vol.~20, no.~10, pp. 1431--1440, 2019.

\bibitem{choe2019deep}
J.~Choe, S.~M. Lee, K.-H. Do, G.~Lee, J.-G. Lee, S.~M. Lee, and J.~B. Seo, ``{Deep learning--based image conversion of CT reconstruction kernels improves radiomics reproducibility for pulmonary nodules or masses},'' \emph{Radiology}, vol. 292, no.~2, pp. 365--373, 2019.

\bibitem{selim2021stan}
M.~Selim, J.~Zhang, B.~Fei, G.-Q. Zhang, and J.~Chen, ``{STAN-CT: Standardizing CT image using generative adversarial networks},'' in \emph{AMIA Annual Symposium Proceedings}, vol. 2020, 2021, p. 1100.

\bibitem{selim2021ct}
------, ``{CT image harmonization for enhancing radiomics studies},'' in \emph{2021 IEEE International Conference on Bioinformatics and Biomedicine (BIBM)}.\hskip 1em plus 0.5em minus 0.4em\relax IEEE, 2021, pp. 1057--1062.

\bibitem{selim2022cross}
M.~Selim, J.~Zhang, B.~Fei, G.-Q. Zhang, G.~Y. Ge, and J.~Chen, ``{Cross-vendor ct image data harmonization using cvh-ct},'' in \emph{AMIA Annual Symposium Proceedings}, vol. 2021, 2022, p. 1099.

\bibitem{kim2022multi}
H.~Kim, G.~Oh, J.~B. Seo, H.~J. Hwang, S.~M. Lee, J.~Yun, and J.~C. Ye, ``{Multi-domain CT translation by a routable translation network},'' \emph{Physics in Medicine \& Biology}, vol.~67, no.~21, p. 215002, 2022.

\bibitem{pei2023multi}
C.~Pei, F.~Wu, M.~Yang, L.~Pan, W.~Ding, J.~Dong, L.~Huang, and X.~Zhuang, ``{Multi-source domain adaptation for medical image segmentation},'' \emph{IEEE Transactions on Medical Imaging}, 2023.

\bibitem{selim2022uda}
M.~Selim, J.~Zhang, B.~Fei, M.~Lewis, G.-Q. Zhang, and J.~Chen, ``{UDA-CT: A general framework for ct image standardization},'' in \emph{2022 IEEE International Conference on Bioinformatics and Biomedicine (BIBM)}.\hskip 1em plus 0.5em minus 0.4em\relax IEEE, 2022, pp. 1698--1701.

\bibitem{zhu2017unpaired}
J.-Y. Zhu, T.~Park, P.~Isola, and A.~A. Efros, ``{Unpaired image-to-image translation using cycle-consistent adversarial networks},'' in \emph{Proceedings of the IEEE international conference on computer vision}, 2017, pp. 2223--2232.

\bibitem{choi2020stargan}
Y.~Choi, Y.~Uh, J.~Yoo, and J.-W. Ha, ``{Stargan v2: Diverse image synthesis for multiple domains},'' in \emph{Proceedings of the IEEE/CVF conference on computer vision and pattern recognition}, 2020, pp. 8188--8197.

\bibitem{10194986}
{Liu, Yunfan and Li, Qi and Deng, Qiyao and Sun, Zhenan and Yang, Ming-Hsuan}, ``Gan-based facial attribute manipulation,'' \emph{IEEE Transactions on Pattern Analysis and Machine Intelligence}, vol.~45, no.~12, pp. 14\,590--14\,610, 2023.

\bibitem{komatsu2021translation}
R.~Komatsu and T.~Gonsalves, ``{Translation of Real-World Photographs into Artistic Images via Conditional CycleGAN and StarGAN},'' \emph{SN Computer Science}, vol.~2, no.~6, p. 489, 2021.

\bibitem{liu2021style}
M.~Liu, P.~Maiti, S.~Thomopoulos, A.~Zhu, Y.~Chai, H.~Kim, and N.~Jahanshad, ``{Style transfer using generative adversarial networks for multi-site mri harmonization},'' in \emph{Medical Image Computing and Computer Assisted Intervention--MICCAI 2021: 24th International Conference, Strasbourg, France, September 27--October 1, 2021, Proceedings, Part III 24}.\hskip 1em plus 0.5em minus 0.4em\relax Springer, 2021, pp. 313--322.

\bibitem{pan2020loss}
Z.~Pan, W.~Yu, B.~Wang, H.~Xie, V.~S. Sheng, J.~Lei, and S.~Kwong, ``{Loss functions of generative adversarial networks (GANs): Opportunities and challenges},'' \emph{IEEE Transactions on Emerging Topics in Computational Intelligence}, vol.~4, no.~4, pp. 500--522, 2020.

\bibitem{mayerhoefer2020introduction}
M.~E. Mayerhoefer, A.~Materka, G.~Langs, I.~H{\"a}ggstr{\"o}m, P.~Szczypi{\'n}ski, P.~Gibbs, and G.~Cook, ``{Introduction to radiomics},'' \emph{Journal of Nuclear Medicine}, vol.~61, no.~4, pp. 488--495, 2020.

\bibitem{huang2017arbitrary}
X.~Huang and S.~Belongie, ``{Arbitrary style transfer in real-time with adaptive instance normalization},'' in \emph{Proceedings of the IEEE international conference on computer vision}, 2017, pp. 1501--1510.

\bibitem{zhang2019self}
H.~Zhang, I.~Goodfellow, D.~Metaxas, and A.~Odena, ``{Self-attention generative adversarial networks},'' in \emph{International conference on machine learning}.\hskip 1em plus 0.5em minus 0.4em\relax PMLR, 2019, pp. 7354--7363.

\bibitem{samuel2011lung}
G.~Samuel, ``{The Lung Image Database Consortium (LIDC) and Image Database resource initiative (IDRI): A completed reference database of lung nodules on CT scans},'' \emph{Medical physics}, vol.~38, p.~2, 2011.

\bibitem{heusel2017gans}
M.~Heusel, H.~Ramsauer, T.~Unterthiner, B.~Nessler, and S.~Hochreiter, ``{Gans trained by a two time-scale update rule converge to a local nash equilibrium},'' \emph{Advances in neural information processing systems}, vol.~30, 2017.

\bibitem{szegedy2016rethinking}
C.~Szegedy, V.~Vanhoucke, S.~Ioffe, J.~Shlens, and Z.~Wojna, ``{Rethinking the inception architecture for computer vision},'' in \emph{Proceedings of the IEEE conference on computer vision and pattern recognition}, 2016, pp. 2818--2826.

\bibitem{van2017computational}
J.~J. Van~Griethuysen, A.~Fedorov, C.~Parmar, A.~Hosny, N.~Aucoin, V.~Narayan, R.~G. Beets-Tan, J.-C. Fillion-Robin, S.~Pieper, and H.~J. Aerts, ``{Computational radiomics system to decode the radiographic phenotype},'' \emph{Cancer research}, vol.~77, no.~21, pp. e104--e107, 2017.

\end{thebibliography}
\bibliographystyle{IEEEtran}

\end{document}